\documentclass{epl}
\usepackage{bm}
\usepackage{amssymb}
\usepackage{psfrag}
\shorttitle{Warped space-time for phonons}
\title{Warped space-time for phonons moving in a perfect 
nonrelativistic fluid}
\author{Uwe R. Fischer\inst{1,2} \and Matt Visser
\inst{3}}
\institute{
  \inst{1} Helsinki University of Technology, Low Temperature
Laboratory, P.O. Box 2200, FIN-02015 HUT, Finland\\
  \inst{2}  Leopold-Franzens-Universit\"at Innsbruck, Institut f\"ur 
Theoretische Physik, Technikerstrasse 25, A-6020 Innsbruck, Austria\\
  \inst{3} School of Mathematical and Computing Sciences, 
Victoria University of Wellington, P.O. Box 600, Wellington, New Zealand
}

\begin{document}

\maketitle
\begin{abstract}
We construct a kinematical analogue of superluminal 
travel in the ``warped'' space-times curved by gravitation, in the form of 
``super-phononic'' travel in the curved effective space-times of 
perfect nonrelativistic fluids. 
These warp-field space-times are most easily generated by
considering a solid object that is placed as an obstruction in an
otherwise uniform flow.
No violation of any condition on the positivity of energy
is necessary, because the 
effective curved space-times for the phonons are ruled by the Euler
and continuity equations, and not by the Einstein field equations. 

\end{abstract}
\newcommand{\MF}{{\large{\manual META}\-{\manual FONT}}}
\newcommand{\manual}{rm}        
\newcommand\bs{\char '134 }     
\newcommand{\kbar}{{{\bf --}}\hspace{-5.5pt}$\kappa$}
\newcommand{\be}{\begin{equation}}
\newcommand{\ee}{\end{equation}}
\newcommand{\bea}{\begin{eqnarray}}
\newcommand{\eea}{\end{eqnarray}}
\def\ii{{\hat\imath}}
\def\jj{{\hat\jmath}}
\def\kk{{\hat k}}
\def\lll{{\hat l}} 
\def\tt{{\hat t}}
\def\xx{{\hat x}}
\def\yy{{\hat y}}
\def\zz{{\hat z}}
\def\Iordanskii{Iordanski\v\i}
\def\projection{\hbox{projection}}
\def\d{{\mathrm{d}}}
\def\ie{{\emph{i.e.}}}
\def\etc{{\emph{etc.}}}
\section{Introduction}

The concept of ``warp fields'', or faster than light (FTL)
propagation/travel, is usually relegated into the realm of science
fiction literature.  Taking warp fields more seriously, when
trying to develop physical realizations within the context of Einstein
gravity, one has to face the difficulty that fulfilling the Einstein
equations demands ``exotic matter''; matter violating the null, weak,
strong, and dominant conditions on the positivity of
energy~\cite{warpdrive,Olum,censorship,bridge,Natario}.  If, on the other
hand, one allows for negative energy densities (which occur for
example in the quantum vacuum of the Casimir effect) the energy
densities (and to a lesser degree the total energies~\cite{Broeck})
actually required to construct macroscopic warp drives are
astronomical~\cite{pfenning-ford}.  One way of side-stepping these
problems, and developing a concrete physical model of what a warp
field might look like, is to consider effective space-time theories
originating in condensed matter~\cite{unruh,vissersonic}: A flowing
hydrodynamical background governed by the nonrelativistic Euler and
continuity equations represents a curved space-time for the
quasiparticle excitations moving in the fluid. The role of the speed
of light is played by the speed of sound, and superluminal
travel turns into super-phononic propagation due to the
effective space-time curvature.  The necessity of violating the energy
conditions is no longer given, because the effective pseudo-Riemannian
space created by the laboratory flow in absolute Newtonian space is
determined by the equations of nonrelativistic hydrodynamics, and not
by the Einstein field equations.

Using the curved space-time analog of quasiparticles propagating on a
hydrodynamical background, we study in this paper a particularly
simple and concrete physical realization of a ``warp field'' realized
by a stationary spherical obstruction in a moving perfect
nonrelativistic fluid.  The effective space-time curvature near the
sphere describes the fact that, when travelling between two specified
points, the (absolute) Newtonian laboratory frame travel time for
quasiparticles which move through the ``warped'' region near the
sphere is either reduced or enhanced; as  compared to the time the
quasiparticles would need to propagate between the same two points
through a flat effective space-time, that is, in a homogeneously 
streaming fluid.

One of the important technical issues involved in warp drives and ``effective
FTL'' is the fact that defining FTL in a standard general relativity
context is subtle
and somewhat subject to coordinate artifacts~\cite{censorship,wald}.
A particularly nice feature of the acoustic geometry presented here is
that it provides a very concrete and definite physical model in which
there is little room for confusion due to coordinate ambiguities, 
because in the present situation the background ``reference metric'' is
unambiguous.

\section{Metric and Curvature}

For the vorticity-free flows considered here, the phonons travelling
around the sphere perceive an acoustic space-time with metric 
\cite{unruh,vissersonic}
\be
\d s^2 = - c_s^2 \; \d t^2 + (\d {\bm x} - {\bm v} \; \d t)^2 \,,
\label{PGmetric} 
\ee
and with Riemann curvature tensor components given in terms of the
deformation tensor~\cite{0205139}
\be 
D_{ij} = \frac12 \left(\partial_i v_j + \partial_j v_i \right).
\ee 
The deformation has vanishing trace for the incompressible background
flow we are assuming, Tr\,$ {\bm D} = 0$, and furthermore $c_s$ is a
constant.  The deformation tensor corresponds in the language of
general relativity to the extrinsic curvature tensor 
\be
K_{ij} =\frac{D_{ij}}{c_s}.
\ee 
In the original proposal by
Alcubierre~\cite{warpdrive}, the extrinsic curvature tensor had
nonvanishing trace, corresponding to a volume element deformation such
that the space contracts in front of the spaceship and expands behind
it. It was, however, recently shown by Nat\'ario~\cite{Natario} that
Tr\,$ {\bm K}$=Tr\,${\bm D}/c_s \neq 0$ is {\em not} a necessary
prerequisite for the warp drive, so that the assumption of an
incompressible background does not impede its construction.

The nonvanishing components of the effective space-time Riemann tensor 
are (for an incompressible, vorticity-free background flow)~\cite{0205139},  
\begin{eqnarray}
\label{E:Riemann1F}
R_{\ii\jj\kk\lll} &=& K_{ik} K_{jl}-K_{il} K_{jk} \,,
\\
\label{E:Riemann3F}
R_{\tt\ii\tt\jj} &=& -\frac{\d}{\d t} K_{ij}-\left({\bm K}^2\right)_{ij}\,.
\end{eqnarray}
The hats indicate that the components are in given in an orthonormal
tetrad basis of the acoustic space-time, and $d/dt=\partial/\partial t
+{\bm v}\cdot \nabla$ is a convective derivative.  Note that if the
flow is steady, the above formulae imply that the Riemann tensor
scales with the flow velocity squared.  The fact that the Riemann
tensor goes to zero quadratically with the flow velocity tells us that
in a theory linearized in the flow velocity, an irrotational,
incompressible fluid flow leads to an intrinsically flat effective
geometry.  Phrased in more conventional language, there are no
(relative acceleration) forces on the quasiparticles~\cite{MeandMatt},
which are linear in the velocity in such a steady flow. 

The Ricci tensor is given by
\begin{eqnarray} 
R_{\tt\tt} 
&=&
- {\rm Tr}({\bm K}^2) 
\,,\qquad 
R_{\ii\jj}  =
\frac{\d}{\d t} K_{ij}\,.
\label{Ricci-t}
\end{eqnarray} 
while the Ricci scalar takes on a particularly simple form
\be
R 
= 
{\rm Tr}({\bm K}^2)\,. 
\label{R}
\ee
Finally the Einstein tensor takes the form
\begin{eqnarray} 
G_{\tt\tt} 
&=&
- \frac12 {\rm Tr}({\bm K}^2) 
\,,\qquad 
G_{\ii\jj}  
=
\frac{\d}{\d t} K_{ij} -\frac12 \delta_{ij}  {\rm Tr}({\bm K}^2) \,.
\label{Einstein-t}
\end{eqnarray} 
Note that $G_{tt} < 0$. This is a purely geometrical statement, which, 
\emph{if} we were to impose the Einstein equations 
$G_{\mu\nu} = 8\pi T_{\mu\nu}$, would immediately
lead to energy condition violations. Because we are not interpreting
the metric in a general relativistic context, we do not use the
Einstein equations. In the acoustic analogue, even though $G_{tt} < 0$
for purely geometrical reasons, energy condition
violations are not implied.

\smallskip
\begin{figure}[htbp]
\psfrag{A}{\Huge ${\bm x}_1,t_1$}
\psfrag{B}{\Huge ${\bm x}_2,t_2$}
\psfrag{x0}{\Huge $x_0$}
\vbox{
\hfil
\scalebox{0.33}{\includegraphics{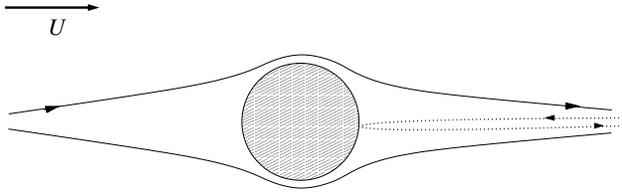}}
\hfil
}
\caption{ \label{sphere}
An impenetrable sphere placed in a stream with velocity ${\bm v}_\infty 
= v_\infty \hat {\bm x}= U c_s \hat{\bm x}$ at infinity. 
Quasiparticles travelling in the flow 
encounter a region of increased effective
space-time curvature near the sphere. 
The dotted line is the reflected phonon trajectory
along the $x$ axis described in the text. 
}
\label{Fig1}
\end{figure}

Consider now the three-dimensional streaming motion of a perfect
liquid past a sphere of radius $a$ (Fig.\ref{sphere}).  Orienting the
co-ordinate system such that the flow velocity at infinity ${\bm
  v}_\infty=v_\infty {\bm e}_x$ is in the $x$ direction, the velocity
components for incompressible irrotational flow are~\cite{llfluids}
\begin{eqnarray}
v_x &=& v_\infty - \frac{ v_\infty a^3}{2r^5}\left( 2x^2 -y^2-z^2 \right) 
=   v_\infty - \frac{ v_\infty a^3}{2r^3}\left( 3 \cos^2 \theta -1 \right) \,,
\nonumber\\
v_y &=&   - \frac{3v_\infty a^3}{2r^5}\, xy 
=  -\frac{3 v_\infty a^3}{4r^3}\, \sin2\theta \cos\phi \,, 
\nonumber\\
v_z &=&  -\frac{3v_\infty a^3}{2r^5}\, xz 
= -\frac{3  v_\infty a^3}{4r^3}\, \sin2\theta \sin\phi \,, 
\label{SphereFlow} 
\end{eqnarray}
where $\theta$ is the angle to the $x$ axis, and $\phi$ is an azimuth.
Note that $||{\bm v}|| \to v_\infty\neq0$ at $r\to\infty$.  These
formulae hold for $r\geq a$. 

To get a qualitative handle on the curvature, note that the velocity
field is of the form
\be 
{\bm v} \sim \hbox{(constant)} + v_\infty {a^3\over r^3} \times \hbox{(direction-dependent factors)}.  
\ee 
This implies that the extrinsic curvature, being given by velocity
gradients, must be of the form
\be 
K \sim U {a^3\over r^4} \times \hbox{(direction-dependent factors)}, 
\ee 
where for convenience we introduce the Mach number $U=v_\infty/c_s$.
Note that our incompressibility assumption forces us to work in the
regime $U\ll1$.  Therefore
\be
\hbox{(Riemann)} \sim U^2 {a^6\over r^8} \times
\hbox{(direction-dependent factors)}.
\label{E:generic}
\ee
Note that $U$ is dimensionless and that the curvature has the correct
dimensions of $1/({\rm length})^2$.

The Ricci curvature scalar (\ref{R}) for the sphere flow
(\ref{SphereFlow}) has the relatively simple anisotropic form
\begin{equation}
R= \frac{9U^2 a^6\left(1+2\cos^2\theta\right)}{2r^{8}}.
\end{equation}
That is, quasiparticles just in front of or behind the sphere
experience the largest space-time curvature.  This anisotropy of the
scalar curvature should be contrasted with the effective space-time
curvature around a vortex~\cite{MeandMatt}, or an impenetrable
infinite cylinder~\cite{0205139}, which both have isotropic $R$,
depending only on the distance from the center of the object placed in
the flow.

\section{Super-phononic propagation}

Defining super-phononic (or in general relativity, superluminal)
propagation in warp field spacetimes is always tricky since by
definition one never permits acoustic propagation outside the local sound
cones of the effective geometry. What one can do, and what is done in
Alcubierre's original analysis \cite{warpdrive} is to compare two
metrics placed on the same manifold. In the presence of the spherical
obstruction we have the metric (\ref{PGmetric}) with ${\bm v}\to{\bm
  v}_\infty$ at spatial infinity. As we have just seen, this metric
leads to spacetime curvature of the acoustic geometry. 
(In general, {\em any} inhomogeneous flow field, which is
asymptotically constant, \emph{i.e.}, yields the above Minkowski form
at infinity, has nonzero curvature.)
If the sphere
were now to be removed, the flow would adjust itself so that the
metric becomes
\be
\d s^2_\infty = - c_s^2 \; \d t^2 + (\d {\bm x} - {\bm v}_\infty  \; \d t)^2
\ee
throughout the spacetime. The spacetime curvature of this acoustic
metric is zero. 
It is by comparing the two metrics $\d s^2$ and $\d s_\infty^2$ that
we can define the notion of super-phononic (superluminal): If the
sphere is absent, the flow is simply ${\bm v}_\infty$ everywhere, the sound
cones are all parallel (they are all tipped over in exactly the same
way) and all have the same opening angle.  Now introduce the sphere
--- it is an obstruction which distorts the flow. The sound cones now
point in different directions at different points in the spacetime.

\section{Time advance: reflection}

For a particularly simple case, think of a phonon that propagates
upstream against the flow and the along the positive $x$ axis from $x_1$ 
to $x_0<x_1$. In the flat reference metric this requires time
$T_0 = {x_1 - x_0 \over c_s - v_\infty}.$
In the curved spacetime in the presence of the sphere, 
sending $x_1\rightarrow \infty$ and $x_0\rightarrow a$, the travel
time becomes  
\be T = 
\int_{a}^{\infty} {\d x\over c_s - v_x(x)} = 
\int_{a}^{\infty} {\d x\over c_s - v_\infty +  v_\infty a^3 / x^3 }.  
\ee 
While the integrals for $T_0$ and $T$ individually diverge, the time
advance, defined as the difference $\Delta T \equiv T-T_0 $, is finite:
\begin{eqnarray} 
\Delta T_{\rm up}
& = &-{a\over c_s} {U\over(1-U)^2} 
\int_1^\infty {\d z \over z^3 + {U\over1-U}}
=-{a\over c_s} {U\over(1-U)^2} 
\sum_{n=0}^\infty {1\over3n+2} \left({-U\over1-U}\right)^n 
\qquad
\nonumber\\
&=&  -{a\over 2c_s} \left\{ U + \frac85 U^2 + O(U^3) \right\}.
\end{eqnarray}
In view of our incompressibility assumption for the background flow, 
keeping higher-order terms in $U$ is meaningless.

The decrease in travel time is the acoustic analog of a ``Shapiro
time advance'' due to the (position-dependent) tipping over of sound
cones, which ultimately connects back to the presence of nontrivial
spacetime curvature. If we consider the surface of the sphere to be a
reflector of sound rays, then upon impact and reflection the sound
waves will be retarded on their downstream journey back to $r=\infty$.
This retardation will not quite cancel the time advance from the
upstream portion of the journey since
\be 
\Delta T_{\rm down} =  \Delta T_{\rm up}(U\to-U).
\ee
The net time advance is 
\be 
\Delta T_{\rm up+down} 
 = - \frac85 {a\over c_s} U^2 + O(U^4).
\ee

The fact that we are seeing a ``Shapiro time advance'' instead of the
more usual ``Shapiro time delay'' characteristic of realistic sources
in general relativity, is because we are \emph{not} enforcing the
Einstein equations, and more specifically not enforcing any positivity
constraint on the components of the Einstein tensor.

\section{Time advance: penetration}

A second situation where the time advance can easily be calculated, 
distinct from the one depicted in Fig. \ref{Fig1},
is when the sphere is taken to be a thin shell, rigid but acoustically
penetrable, and filled with fluid. We now follow a photon upstream from
$\infty$ to $-\infty$. The path divides into three zones:
\begin{itemize}
\item
{From} $\infty$ to $a$, with time advance $\Delta T_{\rm up}$ as previously
calculated.
\item
{From} $a$ to $-a$: through the shell and across the 'quiet zone' inside
the sphere, with time advance
\be
\Delta T_{\rm sphere} 
= {2a\over c_s} -  {2a\over c_s -v_\infty} 
= -{2a\over c_s}\; {U\over 1-U}.
\ee
\item
{From} $-a$ to $-\infty$, still an upstream battle, with time advance
equal to the previously calculated $\Delta T_{\rm up}$.
\end{itemize}
The total time advance is then
\bea
\Delta T_{+\infty\to-\infty} &=& 2\Delta T_{\rm up} + \Delta T_{\rm sphere}
\nonumber
\\
&=&  -{a\over c_s} \left\{ 3 U + \frac{18}5 U^2 + O(U^3) \right\}.
\eea

It should be mentioned that the sort of super-phononic propagation
discussed above is in a sense ``trivial'', as it will be present (to
some extent or another) in any acoustic metric which is both
asymptotically Minkowski and has nontrivial fluid flow. In particular
the effect survives, as we have seen, for arbitrarily weak fluid flow
(arbitrarily weak ``warp fields''). Much more radical was Alcubierre's
suggestion (in the context of general relativity) of placing an
observer inside the warp bubble and letting the warp bubble travel in a
superluminal manner.

\section{Strong warp fields}
\def\vereq#1#2{%
 \lower3pt\vbox{%
  \baselineskip1.5pt
  \lineskip1.5pt
  \ialign{$\math#1\hfill##\hfil$\crcr#2\crcr$\sim$\crcr}%
 }%
}%

In our acoustic context ``strong warp fields'' correspond to $U \lesssim 1$ 
and more radically $U\gtrsim1$, i.e. $v_\infty \gtrsim c_s$, so that 
(in the frame where the sphere is at rest) the asymptotic behaviour of the
fluid flow is supersonic. In this situation we can no longer rely on
the incompressible approximation holding for the background flow (at
least, not for any Euler fluid). There are two ways of proceeding:
\begin{itemize}
\item For $U\lesssim1$ we could solve for the background flow using
  the equation~\cite{llfluids}
\be
\nabla^2 \phi = 
{\nabla_i \phi \over c_s} \;   {\nabla_i \phi \over c_s} \;  {\nabla_i \nabla_j \phi}.
\ee
For $U\ll1$ this reduces to the usual incompressible approximation
$\nabla^2\phi=O(U^2)$. For $U\lesssim1$ the background flow is
distorted away from that of equation (\ref{SphereFlow}), but the
qualitative features of the previous discussion will survive.

For $U\gtrsim1$ a shock wave will generically develop; precluding
actual physical construction of such systems.  For strong warp fields we
should thus think of the analog acoustic geometry as a
{\emph{gedankenexperiment}} that helps us understand some of the
subtleties involved with this sort of effective FTL.

\item Alternatively we could search for a physical system that has two
  distinct and well-separated ``sound'' speeds --- an example of this
  behaviour arising for the quasiparticles present in
  the superfluid $^3$He-A~\cite{Volovik}. If the larger of these ``sound'' 
  speeds (where ``sound'' here means any excitation having linear, i.e.
  relativistic, quasiparticle dispersion)  
  is related to bulk compressibility via $c_{\rm high}^2=dp/d\rho$, while the
  lower ``sound'' speed is governed by the excitations we are interested in,
  then there will be a regime in which $v/c_{\rm high} \ll 1$ while
  $v/c_{\rm low} \gtrsim 1$ (in $^3$He-A, $c_{\rm low}/c_{\rm high}
\sim 10^{-3}$) 
  --- this would permit us to simultaneously
  adopt the incompressibility approximation \emph{and} nevertheless
  have ``supersonic'' flow (with respect to ``slow sound''). For the
  example $^3$He-A, this is possible for two-dimensional situations, 
  e.g.\,\,the flow around a cylinder instead of around a sphere, because 
  $c_{\rm low}$ can be the relevant ``speed of sound'' only 
  in situations with planar symmetry \cite{grishapainleve}.

\end{itemize}
Adopting either viewpoint, recall that the sound cones near spatial
infinity are given by
\be
 {\bm v}_{\mathrm{sound}} =  v_\infty \; \hat {\bm x} + c_s \; \hat {\bm n},
\ee 
and note that in this strong-field case the sound cones are tipped
over so far that all sound is inexorably dragged downstream. In
contrast, inside the bubble one is sheltered from this flow, and a
motion that is ``slower than sound'' in terms of the curved spacetime
metric ($\d s^2$) may be ``faster than sound'' in terms of the flat
spacetime metric ($\d s_\infty^2$).  Indeed, an observer at rest with
respect to the sphere is in this strongly warped situation travelling
``faster than sound'' in terms of the flat spacetime metric. The key
step that allows us to make such pronouncements concerning effective
``superluminal/superphononic'' travel is that there are two natural
metrics that can be placed on the same spacetime, and that these two
metrics can then easily be used for comparison purposes.

In general relativity such a ``two metric'' approach to spacetime is
considerably less natural --- nevertheless, in order to make any sense
of effective FTL one seems to be forced (one way or another) into a
``two metric'' interpretation.  One could (as per Alcubierre) define
the two metrics \emph{by fiat}, effectively by agreeing to only
consider a restricted class of spacetime geometries. Alternatively one
can try to develop specific physical models for the ``reference
metric''. In this article we have seen how such a reference metric
naturally arises in fluid acoustics.

One of the key features of the acoustic geometry, and of analog
models in general, is that they tend to inherit the notion of stable
causality from the background geometry~\cite{scharnhorst}.
Specifically, the Newtonian time parameter $t$ is still always timelike
in the acoustic geometry, and this is very much built in at a
fundamental level --- because of this there is never any risk of
developing closed causal curves in the acoustic geometry. We
\emph{always} have $g^{ab} \nabla_a t \nabla_b t = -1/c_s^2$, so as
long as $c_s^2 >0$ we have $\nabla t$ timelike. (And if $c_s^2 <0$ we
do not get closed timelike curves, such a situation corresponds to an
elliptic equation where sound is in a sense ``infinitely damped'', 
corresponding to Euclidean signature of the metric.) In short,
``chronology protection''~\cite{cpc,cpc-review} is automatic for
acoustic geometries and is not contingent on either the Einstein
equations or quantum physics --- it is built in at the foundations.

\section{Discussion}

In this article we have presented a physical implementation of a
version of Alcubierre's ``warp drive spacetime'' in terms of a condensed
matter system for which we have absolute control over all of the
fundamental physics --- we have translated the notion of FTL travel in
general relativistic ``warp fields'' into a very straightforward and
simple model based on acoustic phonons in a moving fluid. Doing so has
let us carry over basic insight from nonrelativistic fluid mechanics
to clarify subtle issues of general relativity; and conversely the
technical machinery of general relativity can be used as an aid to
visualizing the acoustic properties of a moving fluid. This is merely
one aspect of the ``analogue gravity'' programme, wherein a number of
theorists are working on cross-cultural connections between condensed
matter, general relativity, and particle
physics~\cite{unruh,vissersonic,0205139,MeandMatt,Volovik,Schutzhold,%
Chapline,Leonhardt}.

\section{Acknowledgements}

The research of U.R.F. was supported by the ``Improving Human Potential''
Programme of the European Union under grant No. HPRI 1999-CT-00050, 
by the Austrian Science Foundation
FWF, and the ESF Programme ``Cosmology in the Laboratory.''  




\begin{thebibliography}{66}
  
\bibitem{warpdrive} 
M. Alcubierre, 
Class. Quantum Grav. {\bf 11}, L73 (1994).

\bibitem{Olum} 
K.\,D. Olum, 
Phys. Rev. Lett. {\bf 81}, 3567 (1998).

\bibitem{censorship} 
M. Visser, B.\,A. Bassett, and S. Liberati, 
Nucl. Phys. B (Proc. Suppl.) {\bf 88}, 267 (2000).

\bibitem{bridge}
C.\,Clark, W.\,A.~Hiscock, and S.\,L.~Larson, 
Class.\ Quantum\ Grav. {\bf 16}, 3965 (1999).

\bibitem{Natario} J. 
Nat\'ario, 
Class. Quantum Grav. {\bf 19}, 1157 (2002).

\bibitem{Broeck} 
C. Van Den Broeck, 
Class. Quantum Grav. {\bf 16}, 3973 (1999); 
S. Krasnikov, 
arXiv:gr-qc/0207057. 

\bibitem{pfenning-ford} 
M.\,J.~Pfenning and L.\,H.~Ford, 
Class.\ Quantum\ Grav. {\bf 14}, 1743  (1997).

\bibitem{unruh} 
W.\,G. Unruh, 
Phys. Rev. Lett. {\bf 46}, 1351 (1981).

\bibitem{vissersonic} 
M. Visser, 
Class. Quantum Grav. {\bf 15}, 1767 (1998).


\bibitem{wald}
S.~Gao and R.\,M.~Wald, 
Class.\ Quantum\ Grav.\  {\bf 17}, 4999 (2000).


\bibitem{0205139} U.\,R. Fischer and M. Visser, 
Ann. Phys. (N.Y.) {\bf 304}, 22 (2003). 


\bibitem{MeandMatt} U.\,R. Fischer and M. Visser, 
Phys. Rev. Lett. {\bf 88}, 110201 (2002).

\bibitem{llfluids} 
L.\,D. Landau and E.\,M. Lifshitz: 
{\em Fluid Mechanics}, 
Pergamon Press, Second Edition 1987.

\bibitem{Volovik}
G.\,E.~Volovik, 
arXiv:gr-qc/0104046; 
Phys.\ Rep.\  {\bf 351}, 195 (2001). 

\bibitem{grishapainleve} G.\,E. Volovik, 
JETP Lett. {\bf 69}, 705 (1999) [Pis'ma Zh. \'Eksp. Teor. Fiz. {\bf 69}, 
662 (1999)]. 


\bibitem{scharnhorst}
S.~Liberati, S.~Sonego, and M.~Visser, 
Ann. Phys. (N.Y.)\  {\bf 298}, 167 (2002).


\bibitem{cpc}
S.\,W.~Hawking, 
Phys.\ Rev.\ D {\bf 46},  603 (1992).

\bibitem{cpc-review}
M.~Visser, 
arXiv:gr-qc/0204022.



\bibitem{Schutzhold}
R.~Sch\"utzhold and W.\,G.~Unruh, 
Phys.\ Rev.\ D {\bf 66},  044019 (2002).

\bibitem{Chapline}
G.~Chapline, E.~Hohlfeld, R.\,B.~Laughlin, and D.\,I.~Santiago, 
Phil. Mag. B {\bf 81}, 235 (2001).

\bibitem{Leonhardt}
U.~Leonhardt, 
arXiv:gr-qc/0108085; 
Phys. Rev. A {\bf 62}, 012111 (2000).




\end{thebibliography}
\end{document}